\documentclass[sigconf,natbib=false]{acmart}

\copyrightyear{2026}
\acmYear{2026}
\setcopyright{cc}
\setcctype{by}
\acmConference[MSR '26]{23rd International Conference on Mining Software Repositories}{April 13--14, 2026}{Rio de Janeiro, Brazil}
\acmBooktitle{23rd International Conference on Mining Software Repositories (MSR '26), April 13--14, 2026, Rio de Janeiro, Brazil}
\acmDOI{10.1145/3793302.3793617}
\acmISBN{979-8-4007-2474-9/2026/04}

\RequirePackage[
  datamodel=acmdatamodel,
  style=acmnumeric,
  ]{biblatex}

\usepackage{xcolor}
\usepackage{multirow}
\usepackage{xspace}
\usepackage{fancybox}
\usepackage{enumitem}

\newcommand{\hapr}{HAPR\xspace}
\newcommand{\haprs}{HAPRs\xspace}
\newcommand{\hpr}{HPR\xspace}
\newcommand{\hprs}{HPRs\xspace}

\begin{document}

%\title[Agentic and Human Pull Requests: A Testing-Focused Characterization and Comparison]{Agentic and Human Pull Requests:\\A Testing-Focused Characterization and Comparison}

\title{Human-Agent versus Human Pull Requests: A Testing-Focused Characterization and Comparison}

\author{Roberto Milanese}
\email{roberto.milanese@polito.it}
\orcid{0009-0009-8758-753X}
\affiliation{%
    \institution{Politecnico di Torino}
    \city{Turin}
    \country{Italy}
}
\additionalaffiliation{%
    \institution{University of Molise}
    \city{Campobasso}
    \country{Italy}
}

\author{Francesco Salzano}
\email{francesco.salzano@unimol.it}
\orcid{0000-0002-1029-4861}
\affiliation{%
    \institution{University of Molise}
    \city{Campobasso}
    \country{Italy}
}

\author{Angelica Spina}
\email{angelica.spina@unimol.it}
\orcid{0009-0001-6904-6184}
\affiliation{%
    \institution{University of Molise}
    \city{Campobasso}
    \country{Italy}
}

\author{Antonio Vitale}
\authornotemark[1]
\email{antonio.vitale@polito.it}
\orcid{0009-0009-5840-9777}
\affiliation{%
    \institution{Politecnico di Torino}
    \city{Turin}
    \country{Italy}
}

\author{Remo Pareschi}
\email{remo.pareschi@unimol.it}
\orcid{0000-0002-4912-582X}
\affiliation{%
    \institution{University of Molise}
    \city{Campobasso}
    \country{Italy}
}

\author{Fausto Fasano}
\email{fausto.fasano@unimol.it}
\orcid{0000-0003-3736-6383}
\affiliation{%
    \institution{University of Molise}
    \city{Campobasso}
    \country{Italy}
}

\author{Mattia Fazzini}
\email{mfazzini@umn.edu}
\orcid{0000-0002-1412-1546}
\affiliation{%
    \institution{University of Minnesota}
    \city{Minneapolis}
    \country{USA}
}

\renewcommand{\shortauthors}{Milanese et al.}

% %%
% %% The code below is generated by the tool at http://dl.acm.org/ccs.cfm.
% %% Please copy and paste the code instead of the example below.
% %%


% %%
% %% Keywords. The author(s) should pick words that accurately describe
% %% the work being presented. Separate the keywords with commas.

\begin{abstract}
AI-based coding agents are increasingly integrated into software development workflows, collaborating with developers to create pull requests (PRs). Despite their growing adoption, the role of human-agent collaboration in software testing remains poorly understood. This paper presents an empirical study of 6,582 human-agent PRs (\haprs) and 3,122 human PRs (\hprs) from the \textsc{AIDev} dataset. We compare \haprs and \hprs along three dimensions: (i) testing frequency and extent, (ii) types of testing-related changes (code-and-test co-evolution vs. test-focused), and (iii) testing quality, measured by test smells. Our findings reveal that, although the likelihood of including tests is comparable (42.9\% for \haprs vs. 40.0\% for \hprs), \haprs exhibit a larger extent of testing, nearly doubling the test-to-source line ratio found in \hprs. While test-focused task distributions are comparable, \haprs are more likely to add new tests during co-evolution ($OR=1.79$), whereas \hprs prioritize modifying existing tests. Finally, although some test smell categories differ statistically, negligible effect sizes suggest no meaningful differences in quality. These insights provide the first characterization of how human-agent collaboration shapes testing practices.
\end{abstract}

\maketitle

\section{Introduction}

AI-based coding agents are transforming software engineering by enabling human-agent collaboration throughout the development lifecycle~\cite{qiu2025today,codex,caludecode,copilot,devin,cursor}. These agents are increasingly integrated at the pull request (PR) level to perform complex tasks, including feature implementation, debugging, and testing~\cite{li2025}. As agents become core participants in development, characterizing their testing contributions is crucial for ensuring software quality and reliability. However, current research on agent-based testing often focuses on isolated code generation tasks or specific programming languages~\cite{alves2024detecting, joshi2025disrupting}, overlooking organic development contexts where production and test code co-evolve. While broader studies examine agent adoption across diverse development tasks~\cite{sergeyuk2025using, nikolaidis2024comparison, butler2025dear}, they rarely prioritize testing, leaving a fragmented understanding of how human-agent collaboration shapes testing practices.

To address this gap, we present a large-scale empirical study of software testing in PRs featuring collaboration between human developers and AI-based coding agents (\haprs) versus PRs primarily contributed by human developers (\hprs). We structure our study around three Research Questions (RQs):
\begin{itemize}[noitemsep,topsep=2pt,parsep=0pt,partopsep=0pt]
\item[\textbf{RQ\textsubscript{1}}:] \textit{Are there differences between \haprs and \hprs in the frequency and extent of testing contributions?}
\item[\textbf{RQ\textsubscript{2}}:] \textit{How do the development contexts in which testing contributions appear compare between \haprs and \hprs?}
\item[\textbf{RQ\textsubscript{3}}:] \textit{How does test quality, as reflected by test smells, compare between \haprs and \hprs?}
\end{itemize}
To answer these questions, we analyze 6,582 \haprs and 3,122 \hprs from the \textsc{AIDev} dataset~\cite{li2025}, spanning four programming languages (Java, JavaScript, Python, and TypeScript) and five major coding agents (\textsc{Codex}, \textsc{Claude Code}, \textsc{Copilot}, \textsc{Devin}, and \textsc{Cursor}). We characterize these PRs along three interconnected dimensions that provide a holistic view of human-agent collaborative testing. First, we evaluate testing frequency and extent to determine if agents contribute a volume of test code comparable to \hprs. Second, we examine the development contexts, distinguishing between co-evolution and test-focused tasks, to determine whether the collaboration is primarily focused on test additions or if it also extends to more complex maintenance tasks. Third, we assess testing quality via test smells to ensure that the collaborative effort does not introduce low-quality test code that could hinder long-term maintainability.

Our findings suggest that while human-agent testing practices are largely aligned with traditional workflows, the synergy between developers and agents enables more extensive testing contributions without a corresponding drop in quality. These results provide a foundation for developers to optimize agent integration and for researchers to explore the potential of human-agent collaborative testing. To support the validation and extension of our work, we provide a publicly available replication package~\cite{milanese2025}.

\section{Data Preparation}
\label{sec:data_prep}

For our study, we used the \textsc{AIDev} dataset (version 3)~\cite{li2025dataset}, which provides 932,791 \haprs and 6,618 \hprs from 116,211 GitHub repositories. Since the dataset includes \hprs only for repositories with at least 500 stars, we applied the same filter to \haprs to ensure consistent comparison. We also excluded PRs without changes, leaving us with 12,319 \haprs and 6,521 \hprs from 1,473 repositories.

\paragraph{Language Filtering}
We narrowed our scope to four widely used programming languages: Java, JavaScript, Python, and TypeScript. These languages were selected for three reasons: (i) their popularity in industry~\cite{cass2025} and open-source communities~\cite{github2025}; (ii) the availability of standard testing frameworks~\cite{dasilva2019}, which facilitates the identification of test files; and (iii) the availability of test smell detection tools to support our analysis.
We used \texttt{unidiff}~\cite{bordese2023} to analyze the raw PR diffs retrieved from GitHub (e.g., ~\cite{ard}) and identify all changed files. For each PR, we retrieved file contents from the PR head commit or, in the case of removed files, from the parent of the deletion commit. We then used \textsc{Linguist}~\cite{github2025a} to detect the programming language of each file. After this filtering step, our derived \emph{dataset} included 6,582 \haprs and 3,122 \hprs.
%90,064 files in 6,582 HAPRs (53.4\% of the filtered set) and 3,122 HPRs (47.9\%) from 860 repositories (58.4\%).

\paragraph{Test File Identification} We developed language-specific heuristics to distinguish between test and source code files. Unlike most studies~\cite{kochhar2013, trautsch2017}, which primarily check for the ``test'' keyword in filenames, Gonzalez et al.~\cite{gonzalez2017} proposed using content-specific constructs as more reliable indicators. Inspired by this, we manually analyzed a sample of 370 PRs, stratified by language (95\% confidence level, CL, and 5\% margin of error, ME), from our dataset to derive language-specific heuristics. Each PR was independently reviewed by two authors who are PhD students with at least 3 years of experience in software engineering tasks. They identified test markers based on three aspects: (i) \emph{File Paths} to capture directory and naming conventions, (ii) \emph{Import Statements} to identify testing frameworks imported in test files, and (iii) \emph{Code Keywords} to consider syntactic markers such as annotations, methods, and assertions. We implemented the heuristics and evaluated them on a second stratified sample of 370 PRs. Two authors independently labeled it, achieving a Cohen’s kappa of 0.81, indicating almost perfect agreement~\cite{landis1977}, while disagreements were resolved through discussion. The results show high accuracy in test file identification, with a precision of 0.988, a recall of 0.982, and an F-score of 0.985. Qualitative analysis revealed that the few misclassifications occurred in borderline cases. False positives involved source code used for testing other codebases (e.g.,~\cite{afp}), while false negatives were limited to custom tests that lacked standard patterns (e.g.,~\cite{afn}).

\paragraph{Final Dataset} This data preparation process led to a dataset~\cite{milanese2025} containing 6,582 \haprs and 3,122 \hprs. \emph{2,821 out of 6,582} \haprs and \emph{1,248 out of 3,122} \hprs contain test files. The total number of test files is 17,939, and the total number of changed files across all the PRs in the dataset is 64,542.

\begin{table}[t!]
    \centering
    \scriptsize
    \caption{Testing frequency and extent in \haprs and \hprs.}
    \label{tab:rq1}

    \begin{tabular*}{\columnwidth}{@{\extracolsep{\fill}} p{6em}|c|c|c|p{4em}|c|c|c }
        \toprule
        \multirow{2}{*}{Group} & \multicolumn{3}{c|}{Count} & \multicolumn{4}{c}{Ratio} \\
        \cmidrule{2-8}
        & $TPR$ & $TF$ & $TLoC$ & Setting & \rotatebox{90}{$TPR/PR$} & \rotatebox{90}{$TF/F$} & \rotatebox{90}{$TLoC/LoC$} \\
        \midrule
        \multirow{2}{*}{HAPRs} & \multirow{2}{*}{2,821} & \multirow{2}{*}{9,574} & \multirow{2}{*}{1,158,302} & \textit{All PRs} & 0.429 & 0.204 & 0.327 \\
        & & & & \textit{Test PRs} & - & 0.301 & 0.400 \\
        \midrule
        \multirow{2}{*}{HPRs} & \multirow{2}{*}{1,248} & \multirow{2}{*}{8,365} & \multirow{2}{*}{471,409} & \textit{All PRs} & 0.400 & 0.194 & 0.185 \\
        & & & & \textit{Test PRs} & - & 0.255 & 0.232 \\
        \bottomrule
    \end{tabular*}
\end{table}

\section{Empirical Study}

In this section, we present the methodology and findings for our three research questions, exploring the frequency, context, and quality of testing contributions in \haprs and \hprs.

\subsection{RQ\textsubscript{1}: Frequency and Extent of Testing}

In RQ\textsubscript{1}, we compare the contributions appearing in \haprs and \hprs to assess whether human-agent collaboration affects the likelihood of including tests and the extent of testing-related changes.

\paragraph{Methodology} %We organized our analysis in three levels of detail: pull requests, changed files, and changed lines of code. The first level addresses testing frequency by evaluating \emph{how often} test code is included. The other two levels address the extent of testing by measuring \textit{how much} test code is included. We introduced two metrics, \emph{Count} and \emph{Ratio}, to quantify them. At the PR level, we considered the count of test PRs (TPR) and the ratio to the total number of PRs (TPR$/$PR). At the file and lines of code levels, we considered the count of test files (TF) and test lines (TLoC), and their ratios to the total number of changed files and lines (TF$/$F and TLoC$/$LoC, respectively). 
We organized our analysis across three levels: PRs, changed files, and changed lines of code. The PR level captures testing frequency by assessing \emph{how often} test code is included, while the file and line levels capture testing extent by measuring \textit{how much} test code is included. To quantify these aspects, we defined two metrics: \emph{Count} and \emph{Ratio}. At the PR level, we measured the number of test PRs (TPR) and their ratio to all PRs (TPR/PR). At the file and line levels, we measured the number of test files (TF) and test lines of code (TLoC), along with their ratios to total changed files and lines (TF/F and TLoC/LoC).
In the last two cases, we calculated the ratios against both all PRs (\emph{All PRs} setting) and the subset of test PRs (\emph{Test PRs} setting). This allowed us to understand the extent of testing in general and whether it differs between \haprs and \hprs when tests are included. To assess statistical significance, we applied different tests based on the type of data. At the PR level, we used the chi-squared ($\chi^2$) test at a significance level of $\alpha = 0.05$~\cite{fisher1992} to compare the two groups. Then, we used the odds ratio ($OR$) to measure the effect size~\cite{agresti2006}. At the file and lines of code levels, we first computed the ratio of test files and test lines to the total number of changed files and lines for each PR. This resulted in a distribution of values for \haprs and \hprs, rather than aggregate ratios. Since these distributions were non-normal, as confirmed by the Anderson-Darling test~\cite{yazici2007}, we used the Mann-Whitney U test to compare them at a significance level of $\alpha = 0.05$. Then, we used Cliff's delta ($\delta$) to measure the magnitude~\cite{kitchenham2017} and interpreted it according to the thresholds proposed by Vargha and Delaney~\cite{vargha2000}.

\paragraph{Findings} Table~\ref{tab:rq1} summarizes the results. In terms of testing frequency, \haprs include test code slightly more often than \hprs (42.9\% versus 40.0\%). Although this difference is statistically significant ($p = 0.008$), the odds ratio ($OR = 1.13$) is close to one, suggesting that the difference is minimal. When considering the extent of testing, \haprs demonstrate higher ratios at both the file (0.204 versus 0.194) and the line (0.327 versus 0.185) levels. These differences are statistically significant ($p < 0.001$), though the magnitude is negligible ($\delta_{file} = 0.054, \delta_{line} = 0.053$). In contrast, the differences become more evident in the \emph{Test PRs} setting, where \haprs demonstrate a greater extent of testing at both the file (0.301 versus 0.255) and the line (0.400 versus 0.232) levels. These differences are statistically significant ($p < 0.001$) with small effect sizes ($\delta_{file} = 0.147, \delta_{line} = 0.142$). While the line ratio is nearly double for \haprs, similar Cliff's delta values indicate that the likelihood of an \hapr having a higher test ratio than an \hpr remains consistent across the file and line levels.

\hfill

\noindent\fbox{
\parbox{.957\columnwidth}{
    \textbf{Answer to RQ\textsubscript{1}:} \textit{\haprs and \hprs are similarly likely to include test-related changes, indicating comparable testing frequency. However, when tests are present, \haprs tend to include a larger proportion of test-related files and lines of code, reflecting more extensive testing contributions than those observed in \hprs.}
}
}

\subsection{RQ\textsubscript{2}: Type and Context of Testing}

In RQ\textsubscript{2}, we examine the development contexts in which testing contributions appear to understand whether human-agent collaboration influences where testing work is concentrated within PRs.

\paragraph{Methodology} We manually analyzed a statistically significant stratified sample of 339 test \haprs and 294 test \hprs (95\% CL, 5\% ME). We categorized test development contexts following Zaidman et al.~\cite{zaidman2011}: the co-evolution context (\emph{COC}) involves changes to both production and test code, while the test-focused context (\emph{TFC}) includes changes to test code only. We also considered the test evolution tasks proposed by Pinto et al.~\cite{pinto2012} based on the type of change applied to test files: addition (\emph{Add}), modification (\emph{Mod}), and deletion (\emph{Del}). Combining these contexts and tasks yields six distinct categories. Following a double-coding process, two authors independently reviewed and categorized each PR, assigning all applicable labels and resolving any disagreements through discussion. We use the \emph{Count} and \emph{Ratio} metrics to quantify how frequently the contributions in each category are. To assess the statistical significance, we used the chi-squared ($\chi^2$) test~\cite{agresti2006} at a significance level of $\alpha = 0.05$~\cite{fisher1992} to compare the two groups in each scenario. Since we performed multiple comparisons with the same sample, we adjusted the $p$-values using the Holm-Bonferroni correction~\cite{holm1979}. Finally, we used the odds ratio ($OR$) to measure the effect size~\cite{agresti2006}.

\begin{table}[t!]
    \centering
    \scriptsize
    \caption{Testing type metrics in HAPRs and HPRs under co-evolution (COC) and test-focused (TFC) contexts.}
    \label{tab:rq2}
    \begin{tabular*}{\columnwidth}{@{\extracolsep{\fill}} p{4em}c|c|c|c|c|c|c|c|c }
        \toprule
        \multirow{2}{*}{Group} & & \multicolumn{4}{c|}{$COC$} & \multicolumn{4}{c}{$TFC$} \\
        \cmidrule{3-10}
        & & $Add$ & $Mod$ & $Del$ & Total & $Add$ & $Mod$ & $Del$ & Total \\
        \midrule
        %\multicolumn{9}{c}{Count} \\
        %\midrule
        HAPRs & \multirow{2}{*}{\rotatebox{90}{\textit{Count}}} & 232 &  70 &  3 & 269 & 31 & 43 & 3 & 69 \\
         HPRs & & 161 & 128 & 14 & 237 & 14 & 47 & 1 & 59 \\
        \midrule
        %\multicolumn{9}{c}{Ratio} \\
        %\midrule
        HAPRs & \multirow{2}{*}{\rotatebox{90}{\textit{Ratio}}} & 0.684 & 0.206 & 0.009 & 0.794 & 0.091 & 0.127 & 0.009 & 0.204 \\
         HPRs & & 0.548 & 0.435 & 0.048 & 0.806 & 0.048 & 0.160 & 0.003 & 0.201 \\
        \bottomrule
    \end{tabular*}
\end{table}

\paragraph{Findings} Table~\ref{tab:rq2} summarizes the metrics. The distribution of testing contexts is remarkably consistent between \haprs and \hprs. Specifically, test-and-code co-evolution is significantly more frequent than test-focused tasks, with an 80/20 distribution. However, specific testing tasks reveal a different trend. During co-evolution, \haprs are more likely to include new tests (68.4\% versus 54.8\%). This difference is significant ($p = 0.003$), with a moderate effect size ($OR = 1.79$). Conversely, \hprs are significantly more involved in maintaining and refactoring existing tests (43.5\% versus 20.6\% for modifications and 4.8\% versus 0.9\% for deletions). Both differences are statistically significant ($p_{COCMod} < 0.001$, $p_{COCDel} = 0.023$), and the odds ratios indicate substantial and strong effects for modifications and deletions ($OR_{COCMod} = 0.34$, $OR_{COCDel} = 0.18$), respectively. In contrast, no statistically significant differences were observed in test-focused contexts ($p > 0.05$ for all tasks).

\hfill

\noindent\fbox{
\parbox{.957\columnwidth}{
    \textbf{Answer to RQ\textsubscript{2}:} \textit{Testing contributions in \haprs and \hprs appear in similar development contexts, with both predominantly occurring in code-and-test co-evolution rather than test-focused settings. However, within co-evolutionary contexts, \haprs more frequently introduce new tests, whereas \hprs more often involve the modification or deletion of existing test code.}
}
}

\subsection{RQ\textsubscript{3}: Test Smells}
\label{subsec:rq3}

In RQ\textsubscript{3}, we compare the testing quality of test code introduced in \haprs and \hprs, as reflected by the presence of test smells, to assess if human-agent collaboration influences the quality of tests.

\paragraph{Methodology} We analyzed the presence of test smells in both \haprs and \hprs. For each test PR in our dataset, we analyzed test files in the states before and after each PR using \textsc{AromaDr}~\cite{silva2025aromadr}, a language-independent tool capable of detecting ten common test smells reported in Table~\ref{tab:rq3}. To quantify the impact of these changes, we introduced the \emph{Smell Delta} metric, which measures the difference in the total count of test smells between the pre- and post-PR states. This allowed us to assess the extent to which agents and human authors affect the quality of the test code. To assess statistical significance, we compared the distributions of smell deltas for both groups using the Mann-Whitney U test~\cite{kitchenham2017} at a significance level of $\alpha = 0.05$~\cite{fisher1992}. Finally, we used Cliff's delta ($\delta$) to measure the magnitude of the difference~\cite{kitchenham2017} and interpreted it according to the thresholds proposed by Vargha and Delaney~\cite{vargha2000}.

\begin{table}[t!]
    \centering
    \scriptsize
    \caption{Smell Delta (head vs. base) in \haprs and \hprs.}
    \label{tab:rq3}
    \begin{tabular*}{\columnwidth}{@{\extracolsep{\fill}} p{14em}|p{4em}|c|c|c|c }
        \toprule
        Test Smell & Group & Mean $\Delta$ & Min $\Delta$ & Max $\Delta$ & StdDev \\
        \midrule
        \multirow{2}{*}{Assertion Roulette} & HAPRs & 8.043 & -2353 & 1525 & 72.928 \\
                                            &  HPRs  & 7.698 &  -330 & 1232 & 53.333 \\
        \midrule
        \multirow{2}{*}{Conditional Test} & HAPRs & 1.846 &  -459 & 2686 & 52.996 \\
                                          &  HPRs  & 0.534 &   -58 &   64 &  4.459 \\
        \midrule
        \multirow{2}{*}{Duplicate Assert} & HAPRs & 0.388 &  -18 &   86 &  3.149 \\
                                          &  HPRs  & 0.372 &  -60 &   91 &  4.784 \\
        \midrule
        \multirow{2}{*}{Empty Test} & HAPRs & -0.038 & -168 &  33 & 3.309 \\
                                    &  HPRs  &  0.002 &  -32 &  20 & 1.116 \\
        \midrule
        \multirow{2}{*}{Exception Handling} & HAPRs & 0.504 & -188 & 639 & 13.031 \\
                                            &  HPRs  & 0.271 &  -12 &  29 &  1.771 \\
        \midrule
        \multirow{2}{*}{Ignored Test} & HAPRs & 1.607 & -917 & 467 & 23.124 \\
                                      &  HPRs  & 0.153 &   -8 &  66 &  2.441 \\
        \midrule
        \multirow{2}{*}{Magic Number Test} & HAPRs & 1.607 & -917 & 467 & 23.124 \\
                                           &  HPRs  & 1.650 &  -76 & 336 & 13.269 \\
        \midrule
        \multirow{2}{*}{Redundant Print} & HAPRs & 0.062 & -153 &  61 & 3.463 \\
                                         &  HPRs  & 0.063 &   -7 &   9 & 0.623 \\
        \midrule
        \multirow{2}{*}{Sleepy Test} & HAPRs & 0.076 &  -62 &  51 & 1.879 \\
                                     &  HPRs  & 0.008 &  -13 &  10 & 0.682 \\
        \midrule
        \multirow{2}{*}{Unknown Test} & HAPRs & 4.148 & -301 & 7624 & 148.252 \\
                                      &  HPRs  & 1.926 &   -4 &    9 &   2.306 \\
        \bottomrule
    \end{tabular*}
\end{table}

\paragraph{Findings} Table~\ref{tab:rq3} summarizes the smell distribution across the considered categories for both \haprs and \hprs. Across nearly all test smell categories, \haprs show substantially larger standard deviations and more extreme minimum and maximum values compared to \hprs. This suggests that changes derived from human-agent collaboration tend to be less stable and more heterogeneous: while some \haprs introduce few or no smells, others introduce or remove a very large number of smells. For several categories, most notably \textit{Assertion Roulette} and \textit{Magic Number Test}, both \haprs and \hprs show positive mean deltas. This suggests that these smells are structural issues in test writing, rather than being specific to coding agents. When considering statistical significance, a few smell categories (\textit{Assertion Roulette}, \textit{Conditional Test}, \textit{Magic Number Test} and \textit{UnknownTest}) appear statistically different between \haprs and \hprs; however, in all cases, the effect sizes are negligible, indicating that these differences are not practically meaningful.

\hfill

\noindent\fbox{
\parbox{.957\columnwidth}{
    \textbf{Answer to RQ3:} \textit{Test quality, as reflected by test smells, is largely comparable between \haprs and \hprs. Although a few test smell categories show statistically significant differences, the associated effect sizes are negligible, indicating no practically meaningful differences attributable to human-agent collaboration.}
}
}

\section{Threats to Validity}
\noindent\textbf{Construct Validity.} The heuristic-based identification of test files may result in misclassifications. We mitigate this risk by developing the heuristics through extensive manual analysis of a statistically significant sample of PRs.

\noindent\textbf{Internal Validity.} Observed differences between \haprs and \hprs may be influenced by uncontrolled factors, such as developer expertise or the nature of the addressed issues. While these factors cannot be fully controlled in an observational study, this threat is partially mitigated by analyzing real-world PRs that reflect authentic software development practices rather than artificial settings.

\noindent\textbf{External Validity.} The findings of this study may not fully generalize to all programming languages or coding agents. We mitigate this threat by focusing on widely used programming languages and widely adopted coding agents, as well as on real-world GitHub projects with substantial community engagement.

\section{Related Work}

A few studies have focused on the quality of tests generated by coding assistants. Alves et al.~\cite{alves2024detecting} analyzed test smells in Python test code generated by Copilot, assessing the maintainability of generated tests independently of source code changes. Joshi and Band~\cite{joshi2025disrupting} explored the use of Copilot to generate tests for existing projects and compared the generated tests against manually written tests. While these studies provide valuable insights into the characteristics of agent-generated tests, they primarily consider testing as a standalone activity applied to already implemented functionality, rather than as part of an integrated development process.

Beyond test-specific studies, growing work explores coding agents across broader software development tasks. Sergeyuk et al.~\cite{sergeyuk2025using} examined how coding agents are adopted in practice, focusing on developer perceptions, usage patterns, and organizational implications. Nikolaidis et al.~\cite{nikolaidis2024comparison} compared the effectiveness of ChatGPT and Copilot in generating Python code, evaluating quality and correctness. Butler et al.~\cite{butler2025dear} conducted a controlled trial to assess the impact of generative AI coding tools in real workplace settings, highlighting productivity and behavioral effects. These studies do not focus primarily on testing, nor do they analyze how testing is distributed or shaped in collaborative human-agent workflows.

\section{Conclusion}

%This paper presented an empirical comparison of testing contributions in PRS arising from human-agent collaboration (\haprs) and those primarily authored by humans (\hprs). Our analysis of 6,582 \haprs and 3,122 \hprs shows that, while both groups are similarly likely to include testing contributions, \haprs tend to contain more extensive test-related changes once testing is performed. We further observed systematic differences in testing contexts: \haprs more often introduce new tests during code-and-test co-evolution, whereas \hprs more frequently focus on modifying and maintaining existing test code. Despite greater variability in test smell changes in \haprs, differences in test quality, as measured through test smells, are not practically meaningful.

%These findings provide initial empirical evidence that human– agent collaboration can support substantial testing activity. At the same time, the observed differences suggest that human-agent workflows may shape how testing is carried out. Future work should investigate whether human-agent collaboration would benefit from shifting emphasis toward maintaining and evolving existing tests rather than primarily adding new tests. In addition, future studies should assess the effectiveness of tests produced in \haprs using test coverage and fault-detection capability. Together, these directions can help refine the role of coding agents and inform best practices for AI-assisted software development.

This paper compares testing contributions in 6,582 human-agent PRs (\haprs) and 3,122 human PRs (\hprs). Our analysis reveals that while both groups incorporate testing at similar rates, \haprs nearly double the ratio of test code added. However, the nature of these contributions differs: \haprs predominantly introduce new tests during code-and-test co-evolution, whereas \hprs focus more on modifying and maintaining existing test suites. Despite the significantly higher volume of test code in \haprs, we find no meaningful difference in quality as measured by test smells.

These findings have several implications for the future of AI-assisted software engineering. First, they provide empirical evidence that human-agent synergy can significantly scale testing volume while maintaining a level of quality comparable to that of human-authored tests. Second, the systematic shift toward "addition over maintenance" in co-evolutionary contexts highlights a potential risk in collaboration workflows: a bias toward expanding test suites rather than evolving them. This suggests that while agents are effective for generating new tests, human developers must remain vigilant in directing agents toward refactoring and updating legacy tests to prevent suite bloating and ensure long-term maintainability. Future work should investigate whether human-agent collaboration would benefit from shifting emphasis toward maintaining and evolving existing tests rather than primarily adding new tests. In addition, future studies should assess the effectiveness of tests produced in \haprs using test coverage and fault-detection capability. Together, these directions can help refine the role of coding agents and inform best practices for AI-assisted software development.

\begin{acks}
This publication is part of the project PNRR-NGEU, which has received funding from the MUR – DM 118/2023.
\end{acks}

\balance

\printbibliography

\end{document}